\documentclass[twocolumn,prl,aps]{revtex4}
\usepackage{amsmath}
\usepackage{amssymb}
\usepackage{graphicx}
\usepackage{dcolumn}
\usepackage{graphicx}
\usepackage{dcolumn}
\usepackage{bm}

\begin{document}

\preprint{Phys.Rev.Lett.}

\title{Nonlocal transport
near the charge neutrality point in a two-dimensional electron-hole
system }
\author{G.M.Gusev,$^1$ E.B Olshanetsky,$^2$  Z.D.Kvon,$^2$ A.D.Levin$^1$ N.N.Mikhailov,$^2$
 and S.A.Dvoretsky,$^{2}$}

\affiliation{$^1$Instituto de F\'{\i}sica da Universidade de S\~ao
Paulo, 135960-170, S\~ao Paulo, SP, Brazil}
\affiliation{$^2$Institute of Semiconductor Physics, Novosibirsk
630090, Russia}

\date{\today}
\begin{abstract}
Nonlocal resistance is studied in a two-dimensional system with a
simultaneous presence of electrons and holes in a 20 nm HgTe
quantum well. A large nonlocal electric response is found near the
charge neutrality point (CNP) in the presence of a perpendicular
magnetic field. We attribute the observed nonlocality to the edge
state transport via counter propagating chiral modes similar to
the quantum spin Hall effect at zero magnetic field and graphene near
Landau filling factor $\nu=0$ .


\end{abstract}

\maketitle

Topological insulators have an insulating gapped phase in the bulk
and conducting edge modes, which propagate along the sample
periphery \cite{kane, bernevig, hasan, qi, konig, buhmann}. All
two-dimensional topological insulators (2DTI) can be divided into
two classes: the integer quantum Hall effect (QHE) state
\cite{prange} and the quantum spin Hall effect (QSHE) state
\cite{hasan}, where the transport is described by the chiral/helical
edge modes in the presence/absence of magnetic field B respectively.
The chiral modes in QHE are robust to disorder due to magnetic field
induced time reversal (TR) symmetry breaking \cite{halperin} and
propagate over macroscopic distances. A single pair of helical edge
states in quantum spin Hall effect with an opposite spin
polarization  is also expected to be robust to nonmagnetic disorder
due to preservation of TR symmetry \cite{maciejko}. The quantum Hall
effect state can be realized in any 2D metal in a strong
perpendicular magnetic field, while the quantum spin Hall state
exists in 2D systems with strong spin orbit interaction at B=0. The
QSH state was first discovered in HgTe/CdTe quantum wells
\cite{konig,buhmann}.

\begin{figure}[ht!]
\includegraphics[width=8cm,clip=]{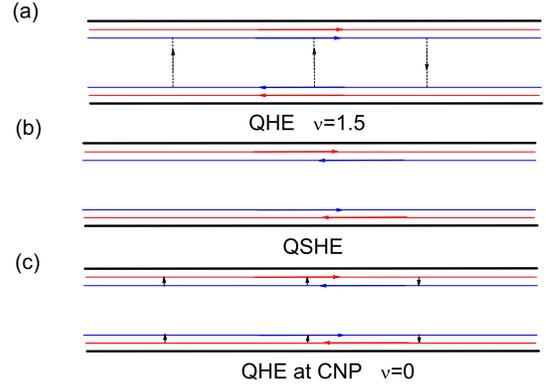}
\caption{\label{fig.1}(Color online) (a) The  chiral edge modes in
the QHE regime at Landau filling factor $\nu=1.5$. Strong
backscattering between the topmost channels  occurs via the bulk
states. (b) The helical edge modes in the QSHE state at B=0. (c)The
counter propagating chiral modes in electron-hole system (and
graphene) in the QHE regime at the charge neutrality point. Small
arrows show the backscattering between edge modes. }
\end{figure}

An unambiguous way to prove the presence of edge state transport
mechanism in a 2DTI are the nonlocal electrical measurements. The
application of the current between a pair of the probes creates a
net current along the sample edge, and can be detected by another
pair of the voltage probes away from the dissipative bulk current
path. Note that the physics of the nonlocality in the QHE regime and
in the QSHE systems is different. Figure 1 illustrates various
transport mechanisms realized in different 2DTIs. In the QHE regime
the nonlocal resistance $R_{NL}$ arises from the suppression of
electron scattering between the outermost edge channels and
backscattering of the innermost channel via the bulk states
\cite{buttiker, mcEuen, dolgopolov}. It may occur only when the
topmost Landau level (LL) is partially occupied i.e at $\nu=n+1/2$,
and the scattering via bulk states is allowed. The transport
measurements \cite{roth, gusev2} in HgTe quantum wells at B=0
reported unique nonlocal conduction properties due to the helical
edge states.

It is worth noting that transport properties of HgTe quantum wells
depend strongly on the well width. As the quantum well width $w$
becomes slightly larger than the "critical" width, approximately
equal to 6.3 nm, the energy spectrum becomes inverted and one gets a
quantum spin Hall insulator. For still higher values of the well
width the quantum well energy spectrum experiences further
transformation. A calculation of the energy spectrum of a wide 20 nm
HgTe QW, has been performed in Ref.[19] taking into account the
strain caused by HgTe/CdTe lattice mismatch. It has been found that
the strain leads to a small overlap of the conductance and valence
band resulting in the formation of a semimetal. The strained 20nm
HgTe QW is the semimetal with the zero gap so it does not have the
quantum spin-Hall effect in contrast to 8nm HgTe samples.  The
transport in such a bipolar system at the charge neutrality point
and in a strong magnetic field is in many respects qualitatively
similar to the quantum Hall effect in graphene. For example, the
resistance was found to increase very strongly with B while the Hall
resistivity turns to zero \cite{gusev}. We attributed the observed
resistance growth to a percolation of the snake-type trajectories.

In this Letter we present an experimental study of the nonlocal
resistance in a 2D bipolar system in undoped 20 nm wide  HgTe
quantum wells with an inverse band structure and (001)
\cite{olshanetsky} surface orientations. We find the nonlocal
resistance increasing significantly with magnetic field near the
charge neutrality point. We explain the observed large nonlocal
resistance using the transport model with counter propagating chiral
edge modes similar to the QSHE at B=0 and graphene at $\nu=0$.

\begin{figure}[ht!]
\includegraphics[width=7cm,clip=]{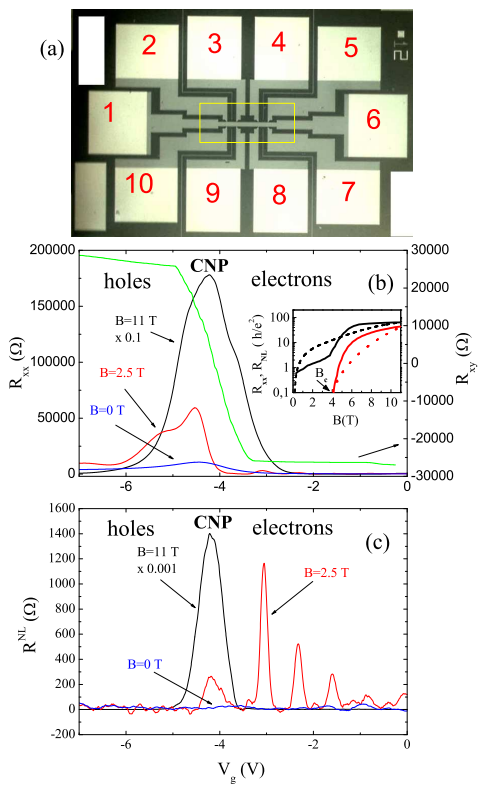}
\caption{\label{fig.2}(Color online) (a) Top view of the sample. The
perimeter of the gate is shown by rectangle. (b) The longitudinal
$R_{xx}$ ( I=1,6; V=3,4) and Hall $R_{xy}$ ( I=1,6; V=3,9)
resistances as a function of the gate voltage at zero and different
nonzero magnetic fields, T=1.4 K. The $R_{xx}$ trace at B=11 T
should be multiplied by 10. (c) The nonlocal $R_{NL}$ ( I=3,9;
V=8,4) resistance as a function of the gate voltage at zero and
different nonzero magnetic fields, T=1,4 K. The trace at B=11 T
should be multiplied by 1000. Insert: Solid traces shows the local (
black) and nonlocal ( red ) magnetoresistance at the CNP as a
function of the magnetic field. Dashed traces show  the local
resistance predicted by the edge+bulk state model.}
\end{figure}

The $Cd_{0.65}Hg_{0.35}Te/HgTe/Cd_{0.65}Hg_{0.35}Te$ quantum wells
with the (001) surface orientations and the width of  20 nm were
prepared by molecular beam epitaxy. A detailed description of the
sample structure has been given in \cite{kvon,gusev, olshanetsky}.
The top view of a typical experimental sample is shown in Figure
2a. The sample consists of three $50 \mu m$ wide consecutive
segments of different length ($100, 250 , 100 \mu m$), and 8 voltage
probes. The ohmic contacts to the two-dimensional gas were formed by
the in-burning of indium. To prepare the gate, a dielectric layer
containing 100 nm $SiO_{2}$ and 200 nm $Si_{3}Ni_{4}$ was first
grown on the structure using the plasmochemical method. Then, the
TiAu gate was deposited. The density variation with gate voltage was
$1.09\times 10^{15} m^{-2}V^{-1}$. The magnetotransport measurements
in the described structures were performed in the temperature range
1.4-70 K and in magnetic fields up to 11 T using a standard four
point circuit with a 3-13 Hz ac current of 1-10 nA through the
sample, which is sufficiently low to avoid the overheating effects.
Several devices from the same wafer have been studied.

Figures 2b and c show the results obtained in the representative
sample. Sweeping the gate voltage $V_{g}$ from 0 to negative values
will depopulate the electron states and populate the hole states. At
$V_{g}=-4.2$ there is a coexistence of electrons and holes with
close densities. The density of the carriers at the CNP without
magnetic field was $n_{s}=p_{s}= 1.2\times10^{10} cm^{-2}$, the
corresponding mobility was $\mu_{n}= 100 000 cm^{2}/Vs$ for
electrons and $p_{s}= 5 000 cm^{2}/Vs$ holes. These parameters were
found from comparison of the Hall and the longitudinal
magnetoresistance traces with the Drude theory for transport in the
presence of two types of carriers \cite{kvon, olshanetsky}. The
local resistance peak $R_{xx}$ corresponding to the CNP increases to
1800 $k\Omega$ at 11 T, whereas the peaks corresponding to higher
Landau levels remain below 10 $k\Omega$. Surprisingly, the nonlocal
resistance $R_{NL}$ measured in configuration (I=3,9;V=4,8), i.e.
where the current flows between contacts 3,9 and the voltage is
measured between contacts 4 and 8 in Figure 2 a, grows from zero at
B=0 to 1400$k\Omega$ a B=11 T and becomes comparable with the local
resistance at the CNP. The peaks in $R_{NL}$ on the electron side of
the CNP in Figure 2 c remain practically the same up to B=11 T and
can be attributed to the well known nonlocality of the quantum Hall
effect edge state transport \cite{mcEuen, dolgopolov}. We don't see
such peaks on the hole side, because the hole mobility is much
smaller than electron mobility. Insert to Fig.2b shows $R_{xx}$ and
$R_{NL}$ at the CNP as a function of the magnetic field. We can see
that the nonlocal magnetoresisance is strongly enhanced above the
critical magnetic field  $B_{c}>4T$. Figure 3 shows the local  (a)
and nonlocal (b) resistance in the voltage-magnetic field plane. One
can see the evolution of both resistances with magnetic field and
density close to the CNP. Nonlocal resistance has a comparable
amplitude, similar peak position, but a narrower width.

The classical ohmic contribution to the nonlocal effect is given by
$R^{classical}_{NL}\approx\ \rho_{xx}\exp(-\pi L/w)$ for narrow
strip geometry where L is the distance between the voltage probes,
and $w$ is the strip width \cite{pauw}. For our geometry we estimate
$R^{classical}_{NL}/R_{xx}\approx 10^{-5}$ for both zero and nonzero
magnetic field. Therefore we can exclude the classical explanation
of the observed nonlocality at finite B while it can possibly
account for the absence of a noticeable nonlocality at B=0. We have
measured the nonlocal response in other geometries, for example,
$R_{NL}$ (I=3,9;V=5,6), and found that the signal is almost
independent of the contact configuration.

We have also examined the local and nonlocal responses near the
CNP as a function of temperature with B fixed at 11 T (Fig.4 a,b).
We see that both resistances increase with the temperature
decreasing. We find that the profile of the $R_{xx}$ and $R_{NL}$
temperature dependencies does not fit the activation law $\sim exp
(\Delta /2kT)$, where $\Delta$ is the activation gap (insert to
Figure 4a), below $T< 10 K$. A similar behaviour has been reported
in our previous study for the local response in the QHE regime at
Landau filling factor $\nu=0$ near the charge neutrality point in
samples with (013) surface orientation \cite{gusev}. Note,
however, that the nonlocal resistance is found to be more
sensitive to temperature than the local resistance: the peak in
$R_{NL}$ disappears completely above 60 K.

Generally a nonlocal response occurs naturally in a 2D system with
the edge state transport. For example in the integer quantum Hall
state all current is carried by the chiral edge states, while
electrons in the bulk region are localized. Note that the bulk
conductivity is shunted by the edge transport and therefore
$R_{NL}\sim h/e^{2}$. This agrees well with our observation of the
nonlocal resistance peaks values in the quantum Hall effect regime
on the electron side of the CNP (figure 2 c) but disagrees with the
behaviour of the nonlocal resistance near the CNP.
\begin{figure}[ht!]
\includegraphics[width=7cm,clip=]{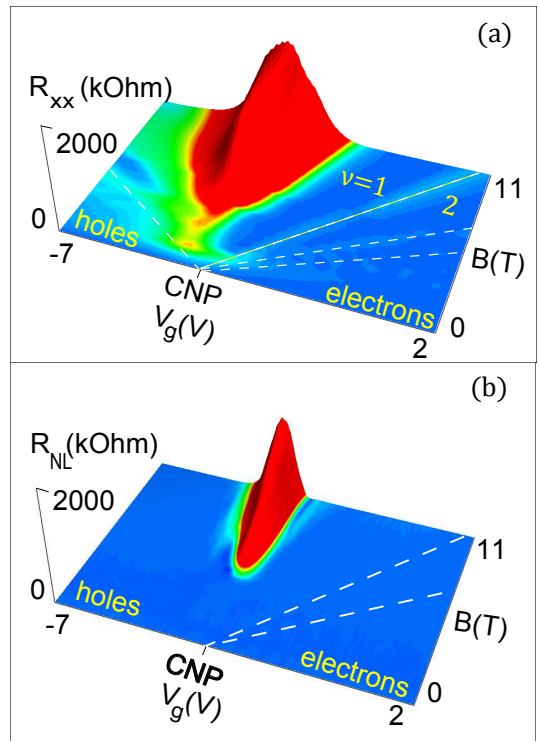}
\caption{\label{fig.3}(Color online)
 The local $R_{xx}$ (a)  and nonlocal (b) resistances as a function
of the gate voltage and magnetic field, T=1.4 K. Dashed lines are
guides to the eye to show the evolution of the resistance peaks.}
\end{figure}
Another example of the edge state transport is the 2D topological
insulator. Our samples have a wider width of 20 nm and  demonstrate
properties of the semimetal rather than topological insulator. The
zero magnetic field behavior of our 20 nm QWs differs from that of 8
nm wide HgTe quantum wells: the 2DTI in an 8 nm QW shows a large
nonlocality, while in the 2D semimetal the nonlocal response is zero
(figure 2c). Application of magnetic field induces a transition from
the helical to the chiral edge states and so suppresses nonlocality
in the 2DTI \cite{gusev2}. In the 2D semimetal, on the contrary,
magnetic field may result in an opposite effect - an enhancement of
the nonlocal resistance due to the edge state contribution to
transport.

\begin{figure}[ht!]
\includegraphics[width=7cm,clip=]{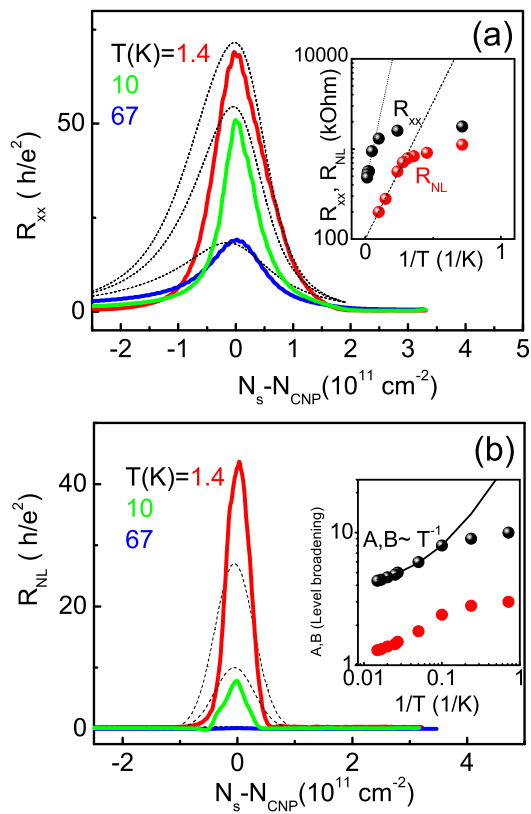}
\caption{\label{fig.4}(Color online) (a) The local resistance
$R_{xx}$ ( I=1,6; V=3,4) as a function of charge density $N_{s}$
swept through the CNP point $N_{CNP}$ at different temperatures,
B=11 T (T(K): 1.4, 10, 67). Dashed traces show $R_{xx}$ predicted by
the model (see \cite{supplementary} and text). Parameter values for
T=1.4 K: A=10, B=3, $\gamma^{-1}=1.42 \mu$m and $g^{-1}=1 \mu$m.
Insert: $R_{xx}$ and $R_{NL}$ as a function of $1/T$ at B= 11 T. The
dashed and dotted lines in the insert are fits of the data with
Arrhenius functions where $\Delta=32$ K and 14 K, correspondingly.
(b) The nonlocal resistance $R_{NL}$ ( I=3,9; V=8,4) as a function
of the gate voltage at different temperatures, B=11 T (T(K): 1.4,
10, 67). Dashed traces show  $R_{NL}$ predicted by the model.
Parameter values are the same as for the local resistance. Insert:
Parameters A( black) and B (red) as a function of $1/T$ at B= 11 T.
The solid traces in the insert are fits of the data A and B with the
$T^{-1}$ dependence. }
\end{figure}

In the rest of the Letter we will focus on the explanation of the
giant nonlocal magnetoresistance observed in our 20 nm wide HgTe
quantum wells with 2D semimetal.

The transport in a 2D semimetals in a strong magnetic field is
equivalent to the QHE state near the CNP in graphene within a
spin-first splitting scenario \cite{abanin1}. The magnetic field
creates a pair of gapless counter propagating edge states. The sharp
peak in the local and nonlocal resistivities near CNP with a
coexistence of electrons and holes can be explained in terms of a
model including simultaneously the edge states and bulk transport
and taking into account the backscattering both between the edge
states (described by one single phenomenological parameter $\gamma$)
and a bulk-edge state coupling (described by a phenomenological
parameter $g$).

Fig.4 shows the modeled behaviour of the local and nonlocal
resistance as a function of the density. Transport coefficient
obtained from the edge+bulk transport model \cite{supplementary}
reproduces the several essential features of the experimental
results. In particular the large peak in the local and nonlocal
resistance is mostly due to the edge transport at the CNP. The
suppression of the peaks away from the CNP is due to a short
circuiting of the edge transport by the bulk conductivity. The
experimental peak of the local resistance is wider than predicted.
This discrepancy is not understood: it may reflect a specific
distribution of the Landau level density of states in the tails. In
particular non-Gaussian tails of the density of states of the Landau
levels were found by many groups in QHE experiments \cite{prange}.
As we mentioned above in our previous study we measured quantum Hall
effect in samples with different surface orientation (113) and large
overlap between electron and hole bands \cite{gusev}. We attribute
the resistance growth at the CNP to a percolation of the snake-type
trajectories along $\nu=0$ lines in the bulk. The bulk conductivity
can thus be governed by snake states. Generally it has been argued
that the electrons on the tails of the Gaussian density of states of
the LL are localized \cite{prange} in the quantum Hall regime. In
our simplified model we did not consider the effects of the
localization, however, we may argue here, that snake states are
delocalized and may contribute to the bulk conductivity near the
CNP.

The  temperature dependence of the local and nonlocal resistance at
the CNP is modeled by the thermal  broadening of the Gaussian Landau
level width for electrons $A$ and holes $B$ \cite{supplementary}.
Such broadening is well known from the temperature dependence of the
integer Quantum Hall effect transition. In particular, it has been
observed that $A\sim T^{-1}$ at high temperatures above 10 K
\cite{prange}. Note that the nonlocal resistance is completely
suppressed above 60 K (figure 4 b) due to a thermally excited bulk
conductivity which shunts the edge state transport at high
temperature. The model also reproduces rapid growth of the local and
nonlocal resistances with magnetic field and threshold-like
behaviour of $R_{NL}$ with B shown in the insert to Fig.2b. However,
this model is much too simple to adequately describe the shape of
the magnetoresistance. In particular the local magneoresistance
demonstrates the different regimes with different transport
mechanisms. Further theoretical and experimental work would be
needed in order to distinguish between all these regimes. Model
predicts even stronger nonlocal effect for smaller bulk-edge
coupling constant g and further increase of the local and non-local
resistivity for more intensive backscattering between edges (larger
parameter $\gamma$). Critical magnetic field $B_{c}$ in the
threshold behavior of $R_{NL}$ depends on the Landau level
broadening- for larger parameters A and B nonlocal resistance is
short circuiting, and  $B_{c}$ is shifted to the stronger magnetic
field.

Recently it has been suggested that spin diffusion may give rise
to nonlocal effects due to a large spin diffusion length, which
can be used for separation of the SHE and the Ohmic contribution
\cite{abanin}. Such giant flavor-Hall effect has been predicted
for semimetals and materials with a Dirac-like energy spectrum and
observed in graphene \cite{abanin3, abanin2}. We would like to
emphasize that despite the similarity of our results and those
obtained in graphene \cite{abanin2} there are, nevertheless,
several essential differences. The main difference is that the
nonlocality in graphene has been detected at low magnetic field
and high temperatures, which points to a quasiclassical origin of
this effect. In contrast, the nonlocality in our system is
observed at low temperatures and high magnetic field, i.e. in the
QHE regime. However, we expect that the spin Hall effect mechanism
may probably be valid in a HgTe-based 2D semimetal system of a
mesoscopic size.

In conclusion, we have observed a large nonlocal resistance in a
20 nm HgTe quantum well, containing simultaneously electrons and
holes in the presence of a perpendicular magnetic field at the
CNP. The nonlocal signal measured between nonlocal voltage
contacts separated from the current probes by 250 $\mu m$ is
comparable in magnitude to the local resistance. We compare our
results to a transport model that takes into account the
combination of the edge state and the bulk transport contributions
and the backscattering within one edge as well as bulk-edge
coupling. The model reproduces many of the key features of the
data, in particular the density and temperature dependence of the
local and nonlocal resistivity.

We thank O.E.Raichev for helpful discussions. A financial support of
this work by FAPESP, CNPq (Brazilian agencies), RFBI grant
N12-02-00054-à and RAS programs "Fundamental researches in
nanotechnology and nanomaterials" and "Condensed matter quantum
physics" is acknowledged.

\end{document}